\documentclass[11pt]{article}           
\usepackage{algorithm,algorithmic}
\newcommand{\BBox}{\rule{0.1in}{0.1in}}

\begin{document}
\title{Integer sorting on multicores: some (experiments and) observations}

\author{Alexandros V. Gerbessiotis\thanks{CS Department, New Jersey Institute of Technology, 
         Newark, NJ 07102, USA. Email: alexg@cs.njit.edu}
}
\maketitle
\thispagestyle{empty}


\thispagestyle{empty}

\begin{abstract}
There have been many proposals for sorting integers on multicores/GPUs
that include radix-sort and its variants or other
approaches that exploit specialized hardware features
of a particular multicore architecture.
Comparison-based algorithms have also been used.
Network-based algorithms have also been used with 
primary example  Batcher's bitonic sorting algorithm.
Although such a latter approach is theoretically ''inefficient'', 
if there are few keys to sort, it can lead to better running
times as it has low overhead and is simple to implement.

In this work we perform an experimental study of integer sorting 
on multicore processors using not only multithreading but also 
multiprocessing parallel programming approaches. 
Our implementations work under Open MPI, MulticoreBSP,
and BSPlib.
We have implemented serial and parallel radix-sort 
for various radixes and also some previously
little explored or unexplored variants of 
bitonic-sort and odd-even transposition sort.

We offer our observations on a performance evaluation using the
MBSP model of such algorithm implementations on multiple platforms 
and architectures and multiple programming libraries. 
If we can conclude anything 
is that modeling their performance by taking into consideration 
architecture dependent features such as the structure and 
characteristics of multiple memory hierarchies is
difficult and more often than not unsuccessful or unreliable. 
However  we can still draw some very simple conclusions  using 
traditional architecture independent parallel modeling.
\end{abstract}



\section{Overview}
\label{overview}

There have been many proposals for sorting integers on multicore
machines including GPUs. These include traditional 
distribution-specific algorithms such as radix-sort 
\cite{BLM91,Garber2008,Langr16,Maus11},
or variants and derivative algorithms of it that use 
fewer  rounds of its baseline count-sort implementation
whenever  more information about the range of key values 
is available \cite{Cheng11,Zhong12}. 
Other proposals include algorithms that use specialized 
hardware or software features of a particular multicore 
architecture \cite{Bramas17, Cheng11,Inoue11,Langr16}.  
Comparison-based algorithms have also been used with some 
obvious tweaks: use of regular-sampling based sorting 
\cite{SH} that utilizes sequential (serial) radix-sort 
for local sorting \cite{Dehne99, Dehne12, Dehne17} or 
not \cite{Zaghloul17,BLM91,Cederman08,Cheng11,Inoue11}. 
Network-based algorithms have also been exploited with  a
primary example being Batcher's \cite{Batcher68} bitonic 
sorting algorithm 
\cite{Ionescu97,BLM91,Peters11,Rathi16,Cederman08}. 
Although such a latter approach can be considered
theoretically ''inefficient'', if there are few keys to 
sort, it can lead to better running times as it has low 
overhead and is simple to implement.

In this work we perform an experimental study of integer sorting 
on multicore processors using not only multithreading but also 
multiprocessing parallel programming approaches. 
Our implementations need only recompilation of the source in
most cases to work under Open MPI \cite{openMPI},
MulticoreBSP \cite{Yzelman14}, and a non-multithreading,
multi-processing and out of maintenace library, 
BSPlib \cite{Hill96}. 
We have implemented  plain-vanilla radix-sort 
(serial and parallel) for various radixes and also
previously little explored or unexplored  variants of
bitonic-sort and odd-even transposition sort methods.

We offer our observations on a performance evaluation of such 
algorithm implementations on multiple platforms and architectures 
and multiple programming libraries. If we can conclude anything 
is that modeling their performance by taking into consideration 
architecture dependent features such as the structure and 
characteristics of multiple memory hierarchies is
difficult and more often than not unusable or unreliable. 
However  we can still draw some very simple conclusions  using 
traditional architecture independent parallel modeling under 
L. G. Valiant's BSP model \cite{LGV90} or the augmented 
MBSP model \cite{G15} that has recently been proposed by this 
author.

For example for very small problem sizes (say $n$, the number
of integer keys is lower than about 10,000) a variant of
bitonic sort as proposed for small size (sample) sorting in 
\cite{GS96, GS99a, GS97a} and that we shall call {\tt BTN}, 
indeed outperforms serial or parallel
radix-sort with their more time-consuming setup and
overhead as long as the number of cores or threads used
is relatively low. This has been observed indepependently
by others eg \cite{BLM91}.
What has been quite even more amazing is
that certain variant of odd-even transposition sort that we
shall call {\tt OET} (in other words, unoptimized bubble-sort), 
might be slightly better, if $p$ the number of threads or cores 
is also kept small. 
This did not use to be the case when hardware configurations
involved $p$ processors of a cluster, an SMP machine, or 
a supercomputer.

Moreover, we have observed that assigning multiple threads per 
core is not recommended for CPUs with large number of cores. 
This is indeed the case where access to main memory
(RAM) is required either because of program or 
data complexity, and non-locality.
And if it is used for CPUs with moderate number of cores, 
it should never exceed the hardware supported bound 
(usually two), or be even lower than that.
If the number of cores is kept small, multiple threads
per core can be used as long as problem size is kept small.
For parallel radix-sort of 32-bit integers 
radix-$2^8$ radix-sorting i.e. four rounds of baseline 
count-sort is faster than the alternative radix-$2^{16}$ 
sorting that uses two rounds. 

However, depending on the architecture and its structure 
of its caches (level-1, level-2 and level-3) it is 
possible that radix-$2^{16}$ to outperform radix-$2^8$.
Overall efficiency is dependent on the number of cores if
the degree of parallelism is large. For degree 
of parallelism less than four or eight, efficiency 
can be expressed either in terms of number of cores or 
threads. Naturally the difference (i.e. ratio) of the 
two nevers exceeds a factor of two (threads over cores) 
as beyond a hardware-regulated two ineffiencies and 
significant drop in performance is observed.
In this latter case number of cores rather than number
of threads determines or best describes speedup or 
efficiency.

\section{Related work}
\label{relatedwork}

The experimental work of \cite{BLM91} offers a collection 
of parallel algorithms that have been used unmodified 
or not as a basis for integer multicore or 
parallel sorting. 
One such algorithm has been radix-sort, another one has been
sample-sort \cite{Reischuk85,HC,Reif87} i.e. randomized 
oversampling-based sorting along the lines of \cite{Reif87}:
in order to sort $n$ keys with $p$ processors,  one uses a 
sample of $ps-1$ uniformly at random selected keys where $s$ 
is a random oversampling factor.
After sorting the sample of size $ps-1$, one then identifies
$p-1$ splitters as equidistant keys in the sorted sample.  
Those $p-1$ keys split the input into $p$ sequences
of approximately the same size than can the be sorted
independently; the analysis of \cite{Reif87}
shows how one can choose $s$ so that each one of the
$p$ sequences is of size $O(n/p)$ with high probability.
A third algorithm used in \cite{BLM91} is bitonic sorting.
For sorting $n$ keys (integer or otherwise) the bitonic
sort of \cite{BLM91} employs a $\Theta(\lg^2{(n)})$ stage
bitonic sort. If the input is properly partitioned
$n/p$ or so of the keys are located inside a single
processor. Thus bitonic merging of those keys can be done
entirely within the corresponding processor as observed
in \cite{BLM91}; they alternatively proposed
using linear-time serial merging over the slower 
bitonic merging.

It is worth noting that the bitonic sort of \cite{BLM91}
for small problem sizes outperformed the other sorting
methods as implemented on a Connection Machine CM-2.

This approach of \cite{BLM91} to bitonic sorting has 
been followed since then. More recently  
\cite{Rathi16} is using an implementation drawn 
from \cite{Ionescu97} for 
bitonic sorting on GPUs and CPUs.
Such an implementation resembles the one above where 
for $n/p$ keys local in-core operations are involved.
Alas, overall GPU performance is rather unimpressive: 
a 10x speedup over the CPU implementation
on an NVIDIA GT520, or 17x on a Tesla K40C for 5M
keys, and speedup in the range of 2-3x for fewer
keys.

However neither \cite{BLM91} nor the other 
implementations of bitonic sort cited
\cite{Rathi16,Ionescu97} or to be cited
later in this section considered that bitonic sorting
involving  $n$ keys on $p$ 
processors/ cores/ threads
can utilize a bitonic network of 
$\Theta(\lg^2{(p)})$ stages.
If $p$ is substantially smaller than $n$, then
the savings are obvious compared to a  
$\Theta(\lg^2{(n)})$ stage bitonic sorter utilized in 
\cite{BLM91,Rathi16,Ionescu97} and other works.

Indeed this author \cite{GS96,GS99a} highlighted the
possibility and used $\Theta(\lg^2{(p)})$ stage bitonic 
sorting for sample sorting involving more than $p$
keys in the context of bulk-synchronous parallel 
\cite{LGV90} sorting.
\cite{GS96,GS99a} cite the work of 
\cite{BS78,Knuth73} for first observing that
a $p$-processor bitonic sorter can sort $n$ keys in
$\lg{(p)} (\lg{(p)}+1)/2$ stages (or rounds). The input
would consist of $p$ sequences of length $n/p$ and
``comparison of two keys'' gets replaced by a 
(serial) merging operation that separates the 
$n/p$ smallest from the $n/p$ largest keys. 
At start-up before the bitonic sorting commences 
each one of the $p$ sequences needs to be sorted.
And if the input is a sequence of integer keys
such sorting can utilize a linear time radix-sort
rather than logarithmic time bitonic-based sorting
or merge-sort.
Such a bitonic sorting method has been implemented
in this work. We shall call this {\tt BTN} in the
remainder.

Odd-even transposition sort \cite{Leighton91} is an
unrefined version of bubble-sort that has
been used for sorting in array structured parallel 
architectures
(one-dimensional arrays, two-dimensional meshes, etc).
In such a sort $n$ keys can be sorted by $n$ processors
in $n$ rounds using an oblivious sorting algorithm.
In an odd-indexed round a key at an odd-indexed position
compares itself to the even-indexed key to its immediate
right (index one more) with a swap if the former key
is greater.  Like-wise for an even-indexed round. 
A very simple observation we make is that if the number 
of processors/cores is $p$, then a $p$-round 
odd-even transposition sort can sort $n$ keys by
dealing with $n/p$-element sequences, one such 
sequence per processor.
Using the remark of \cite{GS96,GS99a} referring to
the work of \cite{BS78,Knuth73} the $p$ sequences 
must be sorted before odd-even transposition sort 
commences its execution on the $n$ key input. 
We shall call this {\sc OET} for future references.

In \cite{Dehne99} an algorithm is presented and
implemented for sorting integers  that utilizes  
the deterministic regular sampling algorithm of 
\cite{SH}.
Deterministic regular sampling \cite{SH} works as follows.
Split regularly and evenly $n$ input keys into $p$
sequences of equal size $n/p$ and then
sort the $p$  sequences independently. Pick then 
from each sequence $p-1$ sample (and equidistant) keys 
for a total sample of size $p(p-1)$.
A serial sorting of the $p(p-1)$ sample keys is
then performed  by one of the $p$ processors/cores.
Subsequentaly that processor selects $p-1$
equidistant splitters from the sorted sample,
broadcasts them to the remaining processors, and all
processors can the split their input keys around
the $p-1$ splitters. A routing operation and subsequent
$p$-way merging of the sorted sequences completes the 
process. In \cite{Dehne99}, for integer sorting,
the first step of independently sorting the $p$ 
sequences independently as well as the last step 
of $p$-way merging are replaced and realized
with a simpler linear-time radix-sort.
One can prove \cite{SH} that if $n$ distinct keys are
split  around  the $p-1$ splitters as explained,
then none of the $p$ resulting  sequences will be of 
size more than $2n/p$. For this method to be optimal 
one needs to maintain $n/p > p^2$. \cite{Dehne99} also
discuss the case of $n/p < p^2$.

As a side note, \cite{GS96, GS99a} extend the
notion of deterministic regular sampling of \cite{SH}
to determininistic regular oversampling. 
The random oversampling factor $s$ whose
behavior was analyzed by \cite{Reif87}
and finetuned in the context of bulk-synchronous
parallel sorting by \cite{GV94} can be transformed
into a deterministic regular oversampling factor:
a sample of $p(p-1)s$ keys can then reduce a
$2n/p$ imbalance into an $(1+\epsilon) n/p$ one
with $\epsilon$ depending on $s$. See
\cite{GS96,GS99a} for details.

In \cite{Dehne12,Dehne17} a variation of the
sample sort of \cite{SH} as used in
\cite{Dehne99} is being utilized for GPU sorting 
of arbitrary (not necessarily integer) keys. 
Similarly but differently from \cite{GS96,GS99a} 
a sample of size $ps$ is being used rather than
the $p(p-1)s$ of \cite{GS96,GS99a}. The GPU architecture's
block thread size determines $p$ and other GPU constraints
dictate $s$. Thus effectively the implied oversampling
factor of \cite{GS96,GS99a} becomes 
$s/(p-1)$ in \cite{Dehne12}. 

The work in \cite{Dehne12} was published around the
time of publication of \cite{Peters11}. The latter work
uses bitonic sorting for GPU sorting. Both
sets of authors' conclusions \cite{Peters11,Dehne12} 
agree than bitonic-sorting works better for small 
values of $n$ and either sample sort \cite{Dehne12} 
is better for the larger $n$ that can be afforded 
by a memory bound and memory bandwidth bound GPU or
radix-sort \cite{Peters11}. 
Thus their overall conclusions are in line with 
those of \cite{BLM91}.

The work of \cite{Bramas17} involves sorting using
AVX-512 instructions on Intel's Knights Landing.
The driver algorithm is quicksort. For small problem sizes
the author explores the use of odd-even transposition sort
or bitonic sort, and picks the latter over the former for
vectorization.

This is reminiscent of the work of \cite{Cederman08} that use 
a quick-sort on the GPU called GPU-Quicksort. The thread 
processors perform a single task: determining whether 
a key is smaller or not than a splitter. Then 
a rearranging of the keys is issued following something akin 
to a scan/parallel prefix \cite{Leighton91} operation. Its 
performance is compared to  a radix-sort implementation and
shown to be slightly better but mostly worse than
a Hybridsort approach that uses bucketsort and merge-sort
(whose performance depends on the distribution of the
input keys).

In \cite{Cheng11} an integer sorting algorithm is presented
that splits the input based on the size of a (shared) L2
cache and private L1 caches and utilizing SIMD instructions
to optimize performance. It reads like a BucketSort 
algorithm. It also assumes that key values are no more 
than a given bound $m$, an assumption that can shave off
rounds from an otherwise value oblivious radix-sort.
A Figure~6 in \cite{Cheng11} show that their algorithm
exhibits a speedup of approximately 3  over serial
radix sort (running time of 3sec for 32,000,000 integers,
over approximately 9.5 sec for serial radix-sort).
What the input distribution of the input keys of the
experiment was, was not clear.

The implementation of AA-sort is undertaken in 
\cite{Inoue11}. AA-sort can be thought of as 
a bubble-sort like enhancement of the odd-even 
transposition sort we discussed earlier
and called {\tt OET}. Whereas in OET we have
odd and even phases in AA-sort there is no such
distinction and either a bubble-sort step 
(from left to right) is executed if a gap parameter 
$g$ has a value of 1, or non-adjacent keys are 
bubble-sorted if $g$ is greater than 1. 
(Thus $a[i]$ and $a[i+g]$ are then compared.)
Likewise to OET initially each $n/p$ sequence is 
sorted; in the case of AA-sort, a merge-sort is
used. 
The use of a bubble-sort oriented approach is to
exploit vectorization instructions of the 
specific target platforms: 
PowerPC 970MP and Cell Broadband Engine(BE).
In the implementations \cite{Inoue11} present also 
results for a bitonic sort based implementation.
AA-sort seems to be slightly faster than bitonic sort 
with SIMD enhancements for 16K random integers. 
In the Cell BE, AA-sort outperforms bitonic sort 
for 32M integers (12.2 speedup for the former 
over 7.1 for the latter on 16 CELL BE cores).

The AQsort algorithm of \cite{Langr16} utilizes 
quicksort, a comparison-based sorting algorithm,
and  OpenMP is utilized to provide a 
parallel/multithread version of quicksort. 
Though its discussion in an otherwise integer
sorting oriented works as this one might not make
sense, there are some interesting remarks
made in \cite{Langr16} that are applicable to
this work.  It is observed
that hyperthreading provides no benefit and that
for Intel and AMD CPUs best performance is obtained
for assigning one thread per core. For Intel Phi
and IBM BG/Q two threads per core provide marginally
lower running times even if four threads are hardware
supported.

In \cite{Maus11} radix-sort is discussed in the
context of reducing the number of rounds of
count-sort inside radix-sort by inspecting
key values' most significant bits. A
parallelized radix-sort along those lines
achieves efficiencies of approximately
15-30\% (speed up of 5-10 on 32 cores).
Other conclusions are in line with 
\cite{Langr16} in that memory channels can't
keep up with the work assigned from many
parallel threads.

In \cite{Zaghloul17} a parallel merge-sort is analyzed
and implemented on multicores. The parallelization
of merge-sort is not optimal. An $(n/p)\lg{(n/p)}$
local and independent sorting on $p$ threads/cores 
is followed by a merging  that takes time 
$n + n/2 + \ldots + n/p \approx 2n$ utilizing 
respectively $1,2, \ldots , p/2$ cores.
Thus the overall speedup of this straightforward
approach is bounded by $n\lg{(n)} / (n \lg{(n)}/p +2n)$.
For $n=2000000$ and $p=8$ a bound on speedup is 
about $4$ and for efficiency around $50\%$. Superlinear
speedups obtained for $p=5,6$ \cite{Zaghloul17} and 
small problem sizes are probably due to caching effects.
This work is similar to that of \cite{Zhong12}. 
The latter deals with multisets (n keys taking
only $k$ distinct values). Even on four processing
cores speedups are limited to less than $50\%$ efficiency.

\section{Implementations}
\label{implementations}

In this section we introduce the algorithms
that we implement and analyze their performance using
the
{\em Multi-memory Bulk-Synchronous Parallel (MBSP) } model 
of computation \cite{G15}.
The MBSP is  parameterized by the septuplet
$(p,l,g,m,L,G,M)$ to abstract, computation and memory interactions
among multi-cores.
In addition to the modeling offered by the BSP model \cite{LGV90} 
and abstracted
by the triplet $(p,l,g)$, the collection of core/processor   components
has $m$ alternative memory units distributed in an arbitrary way,
and the size of the ``fast memory'' is $M$ words of information.
The cost of memory unit-related I/O is modeled by the pair $(L,G)$.
$L$ and $G$ are similar to the BSP parameters $l$ and $g$ respectively.
Parameter $G$ expresses the unit transfer time
per word of information and thus reflects the memory-unit
throughput cost
of writing a word of information into a memory unit
(i.e. number of local computational operations as units of time
per word read/written).
Parameter $L$ abstracts the memory-unit access latency time
that reflects access delays contributed mainly but not exclusively by
two factors:
 (a) unit access-related delays that can not be hidden or amortized by
     $G$, 
and
 (b) possible communication-related costs involved in
accessing  a non-local unit as this could require
intraprocessor or interprocessor communication.

Using the MBSP cost modeling generic cache performance will be
abstracted by the pair $(L,G)$. 
Parameter $m$ would be set to $p$
and $M$ will be ignored; we will assume that $M$ is large enough
to accommodate the radix-related information of radix-sort.
Intercore communication will be abstracted by $(p,l,g)$. Since
such communication is done through main memory $g$ would be the
cost of accessing non-cache memory (aka RAM). 
We shall in the remainder ignore
$l$ and $L$ as we will be modeling our  algorithms at a higher level. 
This is possible because in integer-sorting the operations 
performed are primitive and interaction with memory is 
the dominant operation.
Thus the cost model of an algorithm would abstract only cost of
access to the fast memory ($G$) and cost of access to the slow memory
($g$). Then we will use the easy to abstract $g=5G$ to further
simplify our derivations. This is  based on the
rather primitive thinking that  $20ns$ and $100ns$ reflect access times
to a cache (L2 or higher) and main memory respectively thus defining
a ratio of five between them.

{\bf Serial radix-$r$ radix-sort}

A sequential radix-sort (called {\tt SR4}) was implemented and used
for local independent sorting in the
odd-even transposition sort and bitonic sort implementations.
The radix used was $r=256$ i.e. it is a four-round count-sort.
For such an implementation sorting $N$ keys requires four rounds of
a count-sort. In each round of count-sort the input is read twice,
first during the initial count process and last when the output is
to be generated, and the output is finally written. Thus the cost
of such memory accesses is $3Ng$, with $g$ referring to the cost of
accessing the main memory. Moreoever allocation and initialization of
the count array incurs a cost of $2rG$, with $G$ being the cost of accessing
the fast cache memory. We shall ignore this cost that is dominated by other
terms.
During the count operation the count array is
accessed $N$ times and so is during the output operation for a total cost
of $2NG$.
Thus the overall cost of a round is $3Ng+2NG$.
For all four rounds of 32-bit sorting the total cost is given by the following.
\begin{eqnarray*}
\label{Ts}
 T_s (N,g,G,r) &=& \left( 32 / \lg{(r)} \right) \cdot  \left( 3Ng+2NG \right)
\end{eqnarray*}
If $g=5G$ and $r=256$ then
\begin{equation}
\label{TsG}
 T_s (N,G) = 68 N G
\end{equation}

{\bf Parallel radix-$r$ radix-sort}

We shall denote with {\tt PR2} and {\tt PR4} radix
$r=2^{16}$ and $r=2^8$ parallel radix-sort algorithms.
Ignoring some details that are implementation
dependent such as the contribution of counters used in the
serial part and their copies involved in the parallel part,
we recognize a cost $2rpg$ due to a scatter and gather operations 
involved in the parallel part the algorithm. 
If $n$ keys are to be sorted, each processor or core is assigned
roughly $N=n/p$ keys. A $2NG$ is assigned for the same reasons that was
assigned in the serial version. A $3Ng$ of the serial version will
become $4Ng$ to account a communication required before the output array
is formed in a given round of count-sort. 
\begin{eqnarray*}
\label{Tp}
 T_p (N,g,G,p,r) &=& \left( 32 / \lg{(r)} \right) \cdot  \left( 4Ng + 2NG + 2prg \right)
\end{eqnarray*}
If $g=5G$ and $r=256$ then
\begin{equation}
\label{Tp4}
 T_p (n,G,p) = \left( 88n/p + 40\cdot  256 \cdot p \right) G
\end{equation}
If $g=5G$ and $r=256^2$ then
\begin{equation}
\label{Tp2}
 T_p (n,G,p) = \left(  44n/p + 20 \cdot 256^2 \cdot p \right) G
\end{equation}

{\bf Odd-even transposition sort}

We analyze the algorithm previously referred to as {\tt OET}.
If $n$ keys are to be sorted, each processor or core is assigned
roughly $N=n/p$ keys. First the $N$ keys per processor or core are
sorted using a radix $r=256$ radix-sort independently and in parallel
of each other that requires time $T_s (n/p,G)$. 
Then a $p$ round odd-even transposition sort takes place utilizing
$n/p$ sorted sequences as explained earlier for {\tt OET}.
One round of it requires roughly $4Ng$ 
for communication and  merging (two input and one output arrays). 
Thus the overall cost of all $p$
phases of {\tt OET} will be as follows.
\begin{eqnarray*}
\label{To}
 T_o (n,g,G,p) &=& T_s (n/p,G) + p \left( 4n/p  \right)g 
\end{eqnarray*}
If $g=5G$ and $r=256$ then
\begin{equation}
\label{ToG}
 T_o (n,G,p) = \left( 68 n/p  + 20 n  \right) G
\end{equation}

{\bf Bitonic Sort}

We analyze the algorithm previously referred to as {\tt BTN}.
If $n$ keys are to be sorted, each processor or core is assigned
roughly $N=n/p$ keys. First the $N$ keys per processor or core are
sorted using a radix $r=2^8$ radix-sort independently and in parallel
of each other that requires time $T_s (n/p,G)$. 
Then  $\lg{(p)} (\lg{(p)}+1)/2$ stages of a $p$-processor bitonic-sort
are realized as explained in Section~\ref{relatedwork}.
One round of it requires roughly $4Ng$ 
for communication and  merging/comparing 
(two input and one output arrays). 
Thus the overall cost of all 
stages of bitonic sort will be as follows.
\begin{eqnarray*}
\label{Tb}
 T_b (n,g,G,p) &=& T_s (n/p,G) + \left( \lg{(p)} \cdot (\lg{(p)}+1) /2 \right) \cdot \left( 4n/p  \right) g
\end{eqnarray*}
If $g=5G$ and $r=256$ then
\begin{equation}
\label{TbG}
 T_b (n,G,p)  = \left( 68 n/p + \left( 10 n\lg{(p)}(\lg{(p)}+1) \right) / p \right) G
\end{equation}

\section{Experiments}

All algorithms have been implemented in ANSI C. 
The code has been programmed in such a way that can be
recompiled but does not need to be rewritten and 
works with three 
parallel, multiprocessing or multithreaded programming
libraries: OpenMPI \cite{openMPI}, MulticoreBSP \cite{Yzelman14}, 
and BSPlib \cite{Hill96}.  
The latter library was only used on the Intel platform.
The resulting source code that has been used in these experiments 
is publically and currently available through the author's 
web-page \cite{AVG17}.

A 8-processor quad-core AMD Opteron 8384 Scientific Linux 7 workstation with
128GiB of memory has been used for the experiments. 
We refer to it as the AMD platform in the remainder.
The version of OpenMPI available and used was 1.8.4.
A quad-core Intel Xeon E3-1240 3.3Ghz Scientific Linux 7 workstation with
16GiB of memory has also been used for the experiments.
We refer to it as the Intel platform in the remainder.
The version of OpenMPI available and used was 1.8.1. 
We also run some experiments with version 2.1.1 which was marginally
faster for some experiments. However it caused a problem with some of
our runs that we did not have time to fix so we report results based on 1.8.1.
Version 1.2.0 of MulticoreBSP was used on both platforms and version 1.4 of BSPlib
was used.
The source code is  compiled using the native gcc compiler
{\tt gcc version 4.8.5} with optimization options
{\tt -O2} {\tt -mtune=native} and {\tt -march=native}
and using otherwise the default compiler and library installation.
Indicated timing results (wall-clock time in seconds)  in the tables 
to follow are the averages of four experiments.
We used small problem sizes of $8\times 10^6 , 32 \times 10^6 , 128 \times 10^6$
integers. This is the total problem size, not the per processor size. 
We have also run some experiments for smaller problem sizes to determine
the cut-off point where bitonic-sort or odd-even transposition sort are superior
to radix-sort methods.
The input for all algorithms is the same set of random uniformly drawn integers.

For the serial algorithm {\tt SR4} we report in both Table~\ref{Table1} and
Table~\ref{Table2} the corresponding serial execution time in seconds.
For all other algorithms, {\tt PR4, PR2, BTN} and {\tt OET} we report
speedup figures.  For Table~\ref{Table3}, Table~\ref{Table4}, and Table~\ref{Table5}
we report timing results in microseconds.
We offer some of the observations that we consider
important in the context of this experimentation.

\noindent
{\bf Observation 1: Thread size per core.} 
For the AMD platform one thread per core was a requirement
for extracting best performance. This is in accordance to
remarks by \cite{Langr16} and \cite{Maus11}.

\noindent
{\bf Observation 2: Hyperthreading.} For the Intel platform two 
threads per core was a requirement for extracting consistently 
better performance thus deviating from \cite{Langr16}. 
For smaller or larger problem size
one thread or four threads per core improved speedup slightly but 
to the detriment of (thread) efficiency. 
Thus the one thread per core recommendation or observation of
\cite{Langr16} might not be current any more for more recent
architectures.
For the Intel platform, experiments with $p=8$
on a multicore use exclusively two threads per core. 
OpenMPI was able to cope with it, MulticoreBSP consistently did not,
and so was not (obviously) BSPlib.
The last two suffered a  performance drop by 30\% or so.

\noindent
{\bf Observation 3: Libraries.} For the AMD platform, OpenMPI 
had library latency and performed better in larger problem sizes 
than MulticoreBSP. To some degree the same can be said for 
the Intel platform. BSPlib with its multiprocessing only support
but low library overhead was extremely competitive and bettered 
MulticoreBSP almost always. Moreover it was more often 
than not better than OpenMPI, despite its age and non support.

\noindent
{\bf Observation 4: 4-round vs 2-round radix-sort.} 
On the AMD platform {\tt PR4} was superior to the
low overhead {\tt BTN, OET} implementations, something that was 
not surprising.  However in the AMD platform a two-round 
radix-sort on 32-bit integers was  better than a four-round one 
for both libraries used and for large problem sizes. 
In the case of MulticoreBSP this was so for 32M and 128M but
for the case of OpenMPI only for 128M. 
On the Intel platform a four-round was always the winner.

\noindent
{\bf Observation 5: Bitonic vs Odd-even transposition sort.} 
On the AMD platform {\tt BTN} was a clear winner.
On the INTEL platform surprisingly {\tt OET} was better more 
often than not across all three libraries. Only for $p=8$ 
did it marginally lose to {\tt BTN}. And this is despite 
that $\lg{(p)} (\lg{(p)}+1) /2$ was still smaller than $p$. 
It would be quite interesting to compare the two on
GPU platforms. Note also that our version of bitonic or 
odd-even transposition sort  differs from other approaches 
in that it has only $\lg{(p)} (\lg{(p)}+1) /2$ and $p$ 
stages for {\tt BTN} and {\tt OET} rather than
 $\lg{(n)} (\lg{(n)}+1) /2$ and $n$ respectively.

\noindent
{\bf Observation 6:  Bitonic vs Odd-even transposition sort 
threshold.} 
Under the Intel platform we tested both implementations
for smaller problem sizes ranging from 1K ($=10^3$) to 
about 512K.  These results are shown in
Table~\ref{Table3}, Table~\ref{Table4}, and Table~\ref{Table5} with
figures indicating microseconds, i.e. actual timing results rather than
speedup information.
Under OpenMPI {\tt BTN} 
was better through 128K and only for 512K was {\tt OET} 
marginally better.  Both of them got beaten by the
serial four-round radix-sort through around 32K where both started
getting better running times 
(i.e. speedup greater than 1 or even close to 2 for 512K).
Under MulticoreBSP, {\tt BTN} was marginally better for 
the 1K-32K range and {\tt OET} for the 128K-512K range. 
For the 1K-32K both were
marginally better than serial radix-sort {\tt SR4} and beyond 
that their speedup was in the one to two range over {\tt SR4}.
For BSPlib, {\tt OET} was consistently better than {\tt BTN} 
except for $p=8$ and sizes of 32K or more when it was 
marginally (less than 10\%) slower.
It was for the 128K-512K range that speedup figures were 
slightly over 2 for $p=4$ and 128K and $p=2,4,8$ for  512K.
Both {\tt OET} and to a lesser degree {\tt BTN} exhibited
a speed up of around 2 starting with 2K problem size
This raises the question whether {\tt OET} is just good because 
$p$ is consistently small, or there is some promise to 
it for GPU architectures and appropriate implementations.
But it is safe to say that with increasing processor or thread
sizes {\tt BTN} will prevail.

\noindent
{\bf Observation 7: MBSP modeling {\tt SR4} vs {\tt PR4, PR2}.}
We may use equation~\ref{TsG} and equation~\ref{Tp4} to determine
the relative efficiency of a parallel four-round radix-sort.
We have then than
\begin{eqnarray*}
T_s (n,G) / T_p (n/p,G,p) &=& 68nG / \left( 88n/p + 40 \cdot  256 \cdot p \right)G
\end{eqnarray*}
The fraction to the limit goes to approximately $68*p/88 \approx 0.75p$.
Thus for $p=4,8,16$ we should not be anticipating speedups higher
than about $3, 6$ and $12$ respectively for the AMD platform.
Indeed this is the case from Table~\ref{Table1}.
Likewise for the Intel platform.  In the latter case however,
we have a speedup of $3.61$ over the ''predicted'' $3$ for $p=4$.
Thus we restrained in expecting accuracy from such an estimation
given the assumptions and simplifications incorporated in
the formulas. For example it is possible that
the $g=5G$ is not an accurate one for the Intel Platform.
Repeating this analysis for {\tt PR2} if we use
equation~\ref{TsG} and equation~\ref{Tp2} does not
restrict the possible speedup.
Thus we can make the case that the MBSP model can be used
to reason usefully about the results of our experiments.

\noindent
{\bf Observation 8: MBSP modeling {\tt OET} vs {\tt BTN}.}
If we take the ratio of equation~\ref{ToG} and equation~\ref{TbG}
it is approximately
\[
T_o (n,G,p) / T_b (n,G,p) \approx (68+20p) / (68+10 \lg^2{(p)} ).
\]
For the problem sizes of the experiments and $p=2-16$ we observe
that we expect {\tt OET} to be up to about 40\% slower than {\tt BTN}
for $p=2, 4, 8$ but up to 70\% slower for $p=16$.
This is confirmed for the AMD platform AMD by looking at the speedup
data for the two algorithms. For $p=4, 8$ {\tt BTN} is faster by about
10-25\% than {\tt OET}. For $p=16$ however {\tt BTN} is faster by
50-60\% with respect to {\tt OET} under OpenMPI and 30-65\% under
MulticoreBSP thus confirming this empirical finding.
The fact that this is not reproducible on the Intel platform might
confirm that the ratio of $g/G$ there is different.
Thus it indicates something about the potential of MBSP in
modeling behavior: it might be a useful and usable model but
if one tries to use with the intent of achieving accuracy
the results will be mixed:
it is difficult to model or abstract precisely the interactions of 
the underlying
architecture, its memory hierarchies and its core interactions.

\noindent
{\bf Observation 9: MBSP modeling {\tt BTN} vs {\tt SR4}.}
The ratio of equation~\ref{TsG} and equation~\ref{TbG}
is approximately
\[
T_s (n,G) / T_b (n, G,p) \approx \left( 68 \cdot p \right) / 
\left( 68+10\cdot \lg{(p)} \left( \lg{(p)} +1 \right) \right)
\]
On the AMD platform it suggests that for $p=4, 8, 16$ we should
be expecting speedup figures in the range of  $2.2, 2.9$ and 
4 respectively. 
Indeed this was the case as one can derive from Table~\ref{Table1}.
The highest speedup observed for the corresponding
processor/thread sizes was $2.63,  3.09$ and $2.30$ respectively.
Likewise for the Intel platform it suggests that for 
$p=2, 4, 8$ we should be expecting speedup figures in the
range of $1.6 ,  2.2$ and $2.9$ respectively.
The highest speedup observed for the corresponding
processor/thread sizes are $1.74,  2.26$ and $1.62$ respectively
as obtained from Table~\ref{Table2}.

\section{Conclusion}

We presented an experimental study of integer sorting 
on multicore processors using multithreading and
multiprocessing parallel programming approaches for a
code that is transportable and executable using three
parallel/multithreading libraries, Open MPI, 
MulticoreBSP, and BSPlib.
We have implemented plain-vanilla serial and parallel 
radix-sort for various radixes and also some previously
little explored or unexplored variants of 
bitonic-sort and odd-even transposition sort.

We offered a series of observations obtained through
this evaluations organized and grouped in a way that
has not been done before, to our knowledge.
Some of those observations have been made previously,
but some of them might not be valid any more.

Moreover we expressed the performance of our implementations
in the context of the MBSP model \cite{G15}. We showed
how one can use the model to compare the theoretical
performance of the implementations. Several conclusions
drawn through this theoretical comparison are in line
with the experimental results we obtained. This would
suggest that MBSP might have merit in studying the
behavior of multicore and multi-memory hierarchy algorithms
and thus be a useful and usable model.


\newpage
\begin{table}[h!b!p!]\centering
{\begin{tabular}{|l r|r|r|r||r|r|r|}\hline
\multicolumn{8}{|c|}{Speedup on AMD Platform}  \\ \hline
\multicolumn{2}{|c|}{} &
\multicolumn{3}{c|}{OpenMPI}     &  
\multicolumn{3}{c|}{MulticoreBSP}       
\\ \hline
\multicolumn{2}{|c|}{} &
\multicolumn{1}{c|}{$  8M$}         &
\multicolumn{1}{c|}{$  32M$}        &
\multicolumn{1}{c|}{$  128M$}       &
\multicolumn{1}{c|}{$  8M$}         &
\multicolumn{1}{c|}{$  32M$}        &
\multicolumn{1}{c|}{$  128M$}       \\ \hline
sr4&$p=1$  &  0.362 &  1.459 &  7.454   & 0.362 &  1.459 &  7.454   \\ \hline
pr4&$p=4$  &  2.46  &  2.54  &  3.17    & 2.46  &  2.46 & 3.10   \\
pr4&$p=8$  &  4.02  &  4.19  &  5.27    & 4.36  &  4.27 & 5.42   \\
pr4&$p=16$ &  4.70  &  5.78  &  7.55    & 6.03  &  5.67 & 7.09   \\ \hline
pr2&$p=4$  &  1.84  &  2.43  &  2.72    & 2.51  &  3.16 & 4.41   \\
pr2&$p=8$  &  2.58  &  3.87  &  4.82    & 4.20  &  5.32 & 7.63   \\
pr2&$p=16$ &  3.14  &  5.21  &  7.90    & 5.56  &  7.01 & 9.56  \\ \hline
btn&$p=4$  &  2.03  &  2.02  &  2.44    & 2.18  &  2.11 & 2.63  \\
btn&$p=8$  &  2.51  &  2.34  &  2.88    & 2.76  &  2.49 & 3.09  \\
btn&$p=16$ &  2.19  &  1.84  &  2.30    & 2.91  &  1.97 & 2.35 \\ \hline
oet&$p=4$  &  1.81  &  1.82  &  2.21    & 2.11  &  2.04 & 2.61  \\
oet&$p=8$  &  2.06  &  1.93  &  2.41    & 2.54  &  2.36 & 2.91  \\
oet&$p=16$ &  1.36  &  1.18  &  1.46    & 1.78  &  1.47 & 1.76 \\ \hline
\end{tabular}}
\caption{Speedup for PR4,PR2, BTN and OET on AMD platform; Time(sec) for SR4
\label{Table1}}
\end{table}

\newpage
\begin{table}[h!b!p!]\centering
{\begin{tabular}{|l r|r|r|r||r|r|r||r|r|r|}\hline
\multicolumn{11}{|c|}{Speedup on Intel Platform}  \\ \hline
\multicolumn{2}{|c|}{} &
\multicolumn{3}{c|}{OpenMPI}     &  
\multicolumn{3}{c|}{MulticoreBSP}       &
\multicolumn{3}{c|}{BSPlib}  
\\ \hline
\multicolumn{2}{|c|}{} &
\multicolumn{1}{c|}{$  8M$}         &
\multicolumn{1}{c|}{$  32M$}        &
\multicolumn{1}{c|}{$  128M$}       &
\multicolumn{1}{c|}{$  8M$}         &
\multicolumn{1}{c|}{$  32M$}        &
\multicolumn{1}{c|}{$  128M$}       &
\multicolumn{1}{c|}{$  8M$}         &
\multicolumn{1}{c|}{$  32M$}        &
\multicolumn{1}{c|}{$  128M$}       \\ \hline
sr4&$p=1$ &0.074&0.420&2.878&0.074&0.420&2.878&0.074&0.420& 2.878 \\ \hline
pr4&$p=2$ &1.21& 1.62& 1.79&  0.98& 1.36& 1.65& 1.25& 1.73& 1.87  \\
pr4&$p=4$ &1.27& 1.89& 3.07&  1.21& 1.64& 2.70& 1.68& 2.33& 3.61  \\
pr4&$p=8$ &1.64& 2.41& 4.02&  1.19& 1.66& 2.59& 1.64& 2.32& 3.57  \\ \hline
pr2&$p=2$ &0.55& 0.65& 0.45&  0.64& 0.94& 1.57& 0.59& 0.97& 1.67  \\
pr2&$p=4$ &0.66& 0.73& 0.37&  0.80& 1.19& 2.04& 0.64& 1.27& 2.25  \\
pr2&$p=8$ &1.08& 0.71& 0.51&  0.58& 0.87& 1.52& 0.40& 0.86& 1.63  \\ \hline
btn&$p=2$ &1.15& 1.58& 1.74&  1.13& 1.52& 1.71& 1.15& 1.56& 1.71  \\
btn&$p=4$ &0.97& 1.37& 2.24&  0.93& 1.25& 2.07& 1.01& 1.42& 2.26  \\
btn&$p=8$ &0.69& 0.96& 1.62&  0.58& 0.83& 1.33& 0.67& 0.96& 1.55  \\ \hline
oet&$p=2$ &1.23& 1.71& 1.82&  1.17& 1.57& 1.73& 1.23& 1.67& 1.79  \\
oet&$p=4$ &1.04& 1.47& 2.43&  0.94& 1.31& 2.12& 1.05& 1.50& 2.38  \\
oet&$p=8$ &0.56& 0.93& 1.56&  0.55& 0.78& 1.27& 0.64& 0.92& 1.48  \\ \hline
\end{tabular}}
\caption{Speedup for PR4,PR2, BTN and OET on AMD platform; Time(sec) for SR4
\label{Table2}}
\end{table}

\newpage
\begin{table}[h!b!p!]\centering
{\begin{tabular}{|l r|r|r|r|r|r|r||}\hline
\multicolumn{8}{|c|}{Running time in ($\mu$s) on Intel Platform}  \\ \hline
\multicolumn{2}{|c|}{} &
\multicolumn{6}{c|}{OpenMPI}     
\\ \hline
\multicolumn{2}{|c|}{}            &
\multicolumn{1}{c|}{$  1K$}       &
\multicolumn{1}{c|}{$  2K$}       &
\multicolumn{1}{c|}{$  8K$}       &
\multicolumn{1}{c|}{$ 32K$}       &
\multicolumn{1}{c|}{$128K$}       &
\multicolumn{1}{c|}{$512K$}       \\ \hline
sr4&$p=1$ &  10  & 30  & 90  & 320  &  1330 & 6800  \\ \hline
pr4&$p=2$ & 560  & 420 & 640 & 640  &  1320 & 4000  \\ 
pr4&$p=4$ & 600  & 910 & 700 & 790  &  1800 & 2800  \\
pr4&$p=8$ & 790  & 880 & 990 &1070  &  1550 & 4310  \\ \hline
btn&$p=2$ & 40   &  40 &  80 & 250  &  980  & 3600  \\
btn&$p=4$ & 50   &  70 &  90 & 210  & 1190  & 3000  \\
btn&$p=8$ & 90   & 100 & 150 & 300  &  890  & 3700  \\ \hline
oet&$p=2$ & 70   &  60 & 100 & 260  &  900  & 3200  \\
oet&$p=4$ & 100  & 140 & 160 & 290  & 1230  & 3300  \\
oet&$p=8$ & 270  & 260 & 310 & 480  & 1200  & 4200  \\ \hline
\end{tabular}}
\caption{Timing results in ($\mu$s) for OpenMPI on the Intel platform
\label{Table3}}
\end{table}

\newpage
\begin{table}[h!b!p!]\centering
{\begin{tabular}{|l r|r|r|r|r|r|r||}\hline
\multicolumn{8}{|c|}{Running time in ($\mu$s) on Intel Platform}  \\ \hline
\multicolumn{2}{|c|}{} &
\multicolumn{6}{c|}{OpenMPI}     
\\ \hline
\multicolumn{2}{|c|}{}            &
\multicolumn{1}{c|}{$  1K$}       &
\multicolumn{1}{c|}{$  2K$}       &
\multicolumn{1}{c|}{$  8K$}       &
\multicolumn{1}{c|}{$ 32K$}       &
\multicolumn{1}{c|}{$128K$}       &
\multicolumn{1}{c|}{$512K$}       \\ \hline
sr4&$p=1$ &  10  & 30  & 90  & 320  &  1330 & 6800  \\ \hline
pr4&$p=2$ & 160  & 180 & 220 & 490  &  1090 & 3180   \\
pr4&$p=4$ &  80  &  90 & 140 & 210  &   460 & 1690   \\
pr4&$p=8$ & 120  & 150 & 190 & 260  &   500 & 1590   \\ \hline
btn&$p=2$ &  10  &  20 &  80 & 290  &  1020 & 3700   \\
btn&$p=4$ &   9  &  10 &  40 & 150  &   650 & 3580   \\
btn&$p=8$ &  10  &  10 &  50 & 180  &   790 & 5260   \\ \hline
oet&$p=2$ &  20  &  20 &  80 & 280  &   980 & 3440   \\
oet&$p=4$ &  10  &  10 &  40 & 170  &   680 & 3290   \\
oet&$p=8$ &  20  &  30 &  70 & 230  &   910 & 4950   \\ \hline
\end{tabular}}
\caption{Timing results in ($\mu$s) for MulticoreBSP on the Intel platform
\label{Table4}}
\end{table}

\newpage
\begin{table}[h!b!p!]\centering
{\begin{tabular}{|l r|r|r|r|r|r|r||}\hline
\multicolumn{8}{|c|}{Running time in ($\mu$s) on Intel Platform}  \\ \hline
\multicolumn{2}{|c|}{} &
\multicolumn{6}{c|}{OpenMPI}     
\\ \hline
\multicolumn{2}{|c|}{}            &
\multicolumn{1}{c|}{$  1K$}       &
\multicolumn{1}{c|}{$  2K$}       &
\multicolumn{1}{c|}{$  8K$}       &
\multicolumn{1}{c|}{$ 32K$}       &
\multicolumn{1}{c|}{$128K$}       &
\multicolumn{1}{c|}{$512K$}       \\ \hline
sr4&$p=1$ &  10  & 30  & 90  & 320  &  1330 & 6800  \\ \hline
pr4&$p=2$ &  120 & 240 & 370 & 530  &  1050 & 2970   \\
pr4&$p=4$ &   70 & 80  & 190 & 460  &   810 & 1980   \\
pr4&$p=8$ &  420 & 630 & 910 & 970  &  1250 & 2270   \\ \hline
btn&$p=2$ &   30 & 40  &  70 & 230  &   890 & 3630   \\
btn&$p=4$ &   20 & 20  &  50 & 170  &   670 & 3470   \\
btn&$p=8$ &   10 & 20  &  90 & 220  &   820 & 3350   \\ \hline
oet&$p=2$ &   10 & 20  &  60 & 210  &   870 & 3250   \\
oet&$p=4$ &   10 & 10  &  40 & 160  &   670 & 2830   \\ 
oet&$p=8$ &   20 & 30  &  70 & 240  &   950 & 3870   \\ \hline
\end{tabular}}
\caption{Timing results in ($\mu$s) for BSPlib on the Intel platform
\label{Table5}}
\end{table}

\end{document}